\documentclass{article}
\usepackage[utf8]{inputenc}
\usepackage{dsfont}
\usepackage{amssymb}
\usepackage{amsmath}
\usepackage{amsthm}
\usepackage{amsfonts}
\usepackage{mathabx}
\usepackage{enumitem}
\usepackage{hyperref}
\usepackage{graphicx}
\usepackage{tabularx}
\usepackage{caption}
\usepackage{subcaption}
\usepackage[all]{xy}
\usepackage{etoolbox}

\usepackage{qcircuit}
\usepackage{authblk}
\usepackage{hyperref}

\hypersetup{
    colorlinks=true,
    linkcolor=blue
    }
\begin{document}

\title{Towards an in-depth detection of malware using distributed QCNN}

\author{Tony Quertier\thanks{\texttt{tony.quertier@orange.com}} }
\author{Grégoire Barrué\thanks{\texttt{gregoire.barrue@orange.com}}}

\affil{Orange Innovation, Rennes, France}

\maketitle

\date{}

\begin{abstract}
Malware detection is an important topic of current cybersecurity, and Machine Learning appears to be one of the main considered solutions even if certain problems to generalize to new malware remain.  In the aim of exploring the potential of quantum machine learning on this domain, our previous work showed that quantum neural networks do not perform well on image-based malware detection when using a few qubits.
In order to enhance the performances of our quantum algorithms for malware detection using images, without increasing the resources needed in terms of qubits, we implement a new preprocessing of our dataset using Grayscale method, and we couple it with a model composed of five distributed quantum convolutional networks and a scoring function. We get an increase of around 20\% of our results, both on the accuracy of the test and its F1-score. 
\end{abstract}

\section*{Introduction}

Malicious software detection has become an important topic in business, as well as an important area of research due to the ever-increasing number of successful attacks using malware. With the recent advances in Artificial Intelligence (AI), cybersecurity researchers are shifting their attention to Machine Learning (ML) and Deep Learning (DL) methods to improve malicious files detection \cite{Raff2020, Ucci2019}, and they have been incredibly creative in data preprocessing. In fact, this part is essential now that the learning algorithms are already extremely powerful.

In \cite{Anderson2018}, Anderson et al. trained a feature-based malware detection model using a non-optimized LightGBM algorithm, in \cite{Nataraj2011} Nataraj et al. use k-nearest neighbors algorithm on image-based malware whereas Raff et al. \cite{Raff2017} introduced MalConv, a featureless deep learning classifier using a dense neural network processing raw bytes of entire executable files.
Image-based malware detection is a challenge in both classical and quantum computing, but for different reasons. In classical machine learning, one limitation to obtaining results as good as with standard static features \cite{https://doi.org/10.48550/arxiv.2107.11100} is the number of training data. Because malware images are much more complex and less representative than standard images, convolutional networks need a lot of images to extract information. However, computing resources are not a problem for two-channel images of size $64 \times 64$.

In recent years, research into QML has developed very significantly, whether in the area of learning theory \cite{Cerezo_2022, ragone2023unified}, generalisation capabilities \cite{Abbas_2021,Larocca_2021}, and more practical use-cases as earth observation study by ESA\footnote[1]{https://eo4society.esa.int/projects/qc4eo-study/} or drugs discovery \cite{batra2021quantum}. Several models of quantum algorithms were developed recently in the field of QML. In particular, quantum convolutional neural networks (QCNNs) have proved to be interesting, because they perform well, and for example do not exhibit Barren Plateaus \cite{Pesah_2021}. They were first introduced in \cite{cong2019quantum}, and used for instance in  \cite{Hur_2022} for classical data classification, or in \cite{Bokhan_2022} for multi-class classification. Their full understanding is still missing, but some works, as \cite{umeano2023learn} start to deeply analyze these models, and we think that it could be a suitable solution for image classification, as is its classical equivalent.

For our problem, the challenge also comes from the limited number of qubits. On the features, we have restricted the number of qubit to less than 8 \cite{barrue2023quantum}, but for image-based detection this number is not enough. We still want to give an approach using relatively few qubits, to keep coherence with the current capacities of the quantum computers,  so we try to find suitable solutions for extracting more information from our data. 
Due to all these limitations, quantum neural networks fail to achieve satisfying results for malware detection using image transformation. Indeed, it's quite challenging to extract relevant information from a malware image of size 64x64 with just a few qubits. During our previous experiments, we came up against these issues, so we looked for a solution to get around this limitation. First we turn ourselves to quantum convolutional neural networks, which perform better on image classification, and we split each image of binary file into sub-images, corresponding to  specific sections of the binary file. 

In this paper, we present an algorithm consisting of five distributed QCNNs trained on the images corresponding to the different sections of a binary file and a scoring function to extract as much information as possible from it. We obtain very satisfying results for accuracy on a dataset compared with our previous work on this subject.
 
\section{Dataset and preprocessing}

\subsection{Dataset}

For our experiments, we rely on two different datasets, Bodmas \cite{Yang} and PEMachineLearning \cite{Anderson2018}. Bodmas shared with our team a dataset that contains 57, 293 malicious files in raw PE format. These files have been collected during one year between August 2019 and September 2020. The second dataset used is PEMachineLearning, made available by M. Lester \cite{web1}. It contains 201, 549 binary files
including 114, 737 malicious files. More details on these two datasets are available in the paper \cite{marais2022ai}. We split the dataset into three sub-datasets, one to train the QCNNs on each section, one to train the final scoring function and one to test our architecture. Each sub-dataset is composed of 20k benign and 20k malware and the train/validation ratio is $70/30$ for the first two.

\subsection{PE file format}

The PE format is a file format used by Windows operating systems.
A PE file is separated into two parts: the header and the sections. The header describes the file and its contents. It contains information such as the date the file was created, information about loading the file into memory and the number of sections. 
Each section is described by a specific header containing its name, size and location in virtual memory. Sections generally contain the executable code (.text) and the variables used with their default values (.data, or in read only .rdata). The relocation section (.reloc) contains relocation information and the resource section (.rsrc) contains resources like icons, menus, and other elements. The resource section is a very interesting section, as it is often used by malware to evade detection. For example, scripts can be used to inject payloads directly into this section. When the binary file is executed, the embedded payload is extracted and executed.

\subsection{Splitting the image into sub-images}

When preprocessing our data, we are using the Grayscale method, that transforms the malware into an image \cite{Nataraj2011}, but in a more subtle way that is better adapted to our problem. We first explore the PE file using the LIEF library \cite{LIEF} to identify the start and end of each section. Then we transform the content of each section into an image of size $8 \times 8$. To begin with, we have identified 5 sections that we consider to be relevant and we focus on these when they are present. If they are not, we give these sections a score of $-1$. Initially we had chosen a neutral score of $0.5$, but this introduced a bias because this is equivalent to considering that the absence of a section is of no importance.. However, the presence or absence of certain sections can provide additional information about a binary file.

\begin{figure}[!h]
    \centering
    \includegraphics[width=0.65\textwidth]{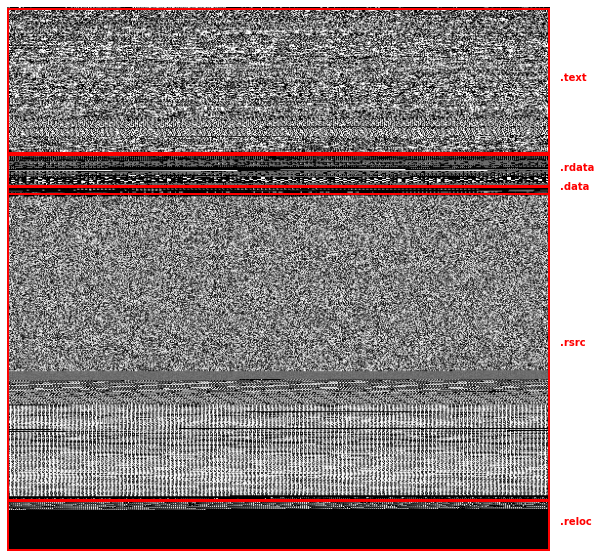}
    \caption{Sections in a malware}
    \label{fig:sections}
\end{figure}

\begin{figure}[!h]
    \centering
    \begin{subfigure}[h]{0.2\textwidth}
        \includegraphics[width=\textwidth]{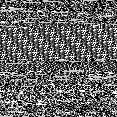}
        \caption{example of a section ".text"}
    \end{subfigure}
    \hspace{1cm}
    \begin{subfigure}[h]{0.2\textwidth}
        \includegraphics[width=\textwidth]{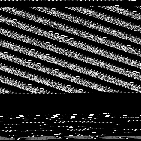}
        \caption{example of a section ".data"}
    \end{subfigure}

        \begin{subfigure}[h]{0.2\textwidth}
        \includegraphics[width=\textwidth]{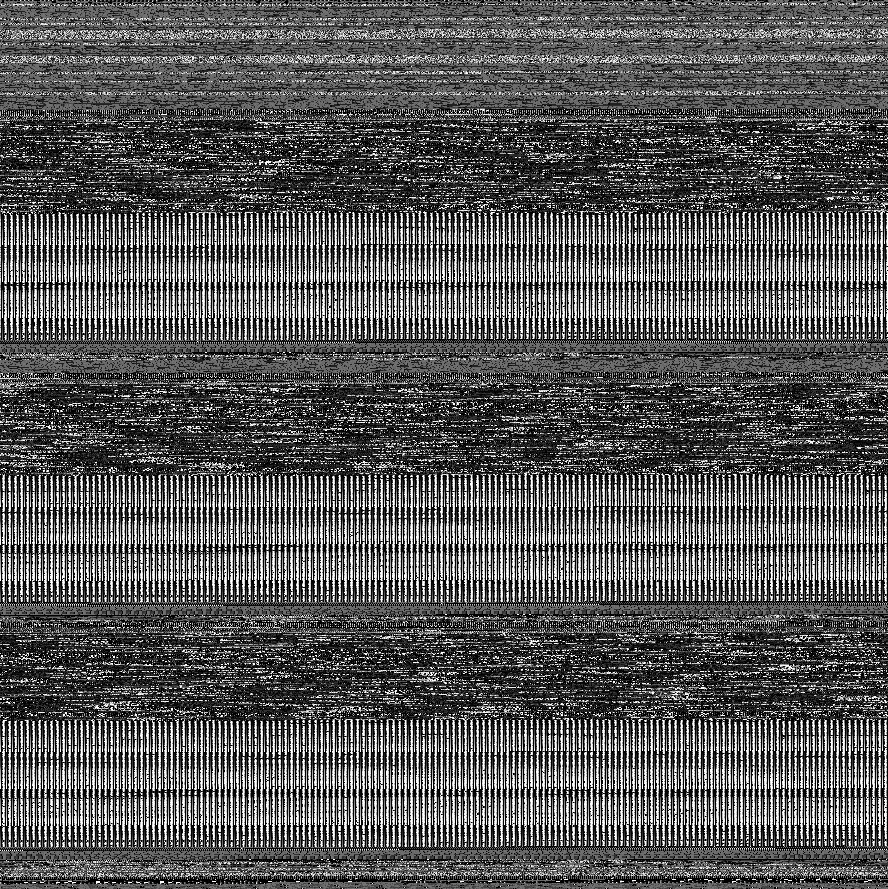}
        \caption{example of a section ".rdata"}
    \end{subfigure}
    \hspace{1cm}
    \begin{subfigure}[h]{0.2\textwidth}
        \includegraphics[width=\textwidth]{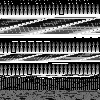}
        \caption{example of a section ".rsrc"}
    \end{subfigure}
    \hspace{1cm}
    \begin{subfigure}[h]{0.2\textwidth}
        \includegraphics[width=\textwidth]{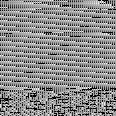}
        \caption{example of a section ".reloc"}
    \end{subfigure}
    \caption{Images of different binary's sections }
\end{figure}

\section{Framework}

\subsection{Description of the QCNN and its training on a section image}

\begin{figure}[!h]
    \centerline{\Qcircuit @C=1em @R=.7em {& \multigate{7}{U_\phi} & \qw & \multigate{1}{C_1}  &\ctrl{1} \\
    & \ghost{U_\phi} & \qw & \ghost{C_1} & \gate{P_1} & \qw & \multigate{2}{C_2} &  \ctrl{2} \\
    & \ghost{U_\phi} & \qw & \multigate{1}{C_1} &  \ctrl{1} \\
    & \ghost{U_\phi} & \qw & \ghost{C_1} &  \gate{P_1} & \qw & \ghost{C_2} & \gate{P_2} & \qw & \multigate{4}{C_3}  & \ctrl{4} \\
    & \ghost{U_\phi} & \qw & \multigate{1}{C_1} &  \ctrl{1} \\
    & \ghost{U_\phi} & \qw & \ghost{C_1} &  \gate{P_1} & \qw & \multigate{2}{C_2} & \ctrl{2} \\
    & \ghost{U_\phi} & \qw & \multigate{1}{C_1} &  \ctrl{1} \\
    & \ghost{U_\phi} & \qw & \ghost{C_1} &  \gate{P_1} & \qw & \ghost{C_2} & \gate{P_2} & \qw & \ghost{C_3} & \gate{P_3} & \qw & \meter & \cw & && \lstick {\langle \sigma_Z\rangle}
    }}
    \caption{The architecture of our QCNN, where convolutional and pooling layers alternate up to the measurement by an observable.}
    \label{fig:QCNN}
\end{figure}
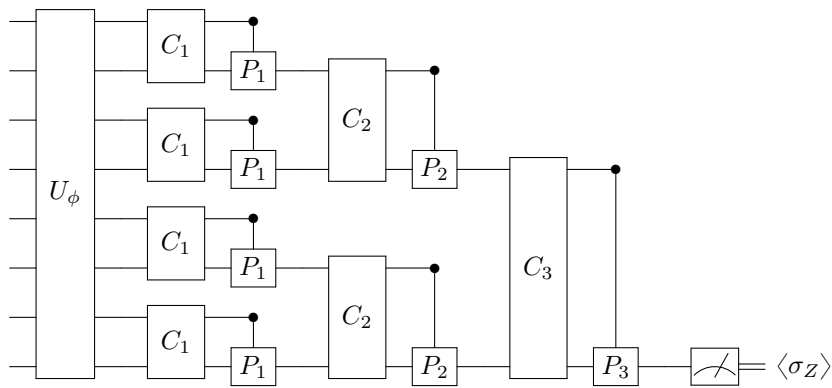

Here we present the architecture of our algorithm. The first step is to train a QCNN for each chosen section. To do so, we start by generating images for each section extracted from the file, then we train the QCNNs on the different sections by setting the number of qubits to 8 and therefore using 3 layers, as shown in Figure \ref{fig:QCNN}.

\begin{figure}[!h]
    \centering
    \begin{subfigure}[!h]{0.2\textwidth}
        \centerline{\Qcircuit @C=1em @R=.7em {& \ctrl{1} & \qw & \ctrlo{1} & \qw & \meter\\
        & \gate{R_Z(\theta_1)} & \qw & \gate{R_X(\theta_2)} & \qw & \qw }}
        \caption{Pooling layer}
        \label{fig:pooling}
    \end{subfigure}
    \hspace*{3cm}
    \begin{subfigure}[!h]{0.3\textwidth}
        \centerline{\Qcircuit @C=1em @R=.7em { & \gate{R_Y(\theta_1)} &\ctrl{1} & \qw \\
         & \gate{R_Y(\theta_2)} & \targ & \qw }}
        \caption{Convolutional layer}
        \label{fig:convolution}
    
    \end{subfigure}
    \caption{Description of the convolutional and pooling layers used in our algorithm. In the pooling layer, the symbol $\bullet$ means that the gate $R_Z(\theta_1)$ is activated only if the first qubit is in the state $|1\rangle$. Conversely, the symbol $\circ$ means that the gate $R_X(\theta_2)$ is activated only if the first qubit is in the state $|0\rangle$. At the end of this subcircuit we trace out the control qubit in order to reduce the dimension.}
    \label{fig:pooling_convolution}
\end{figure}
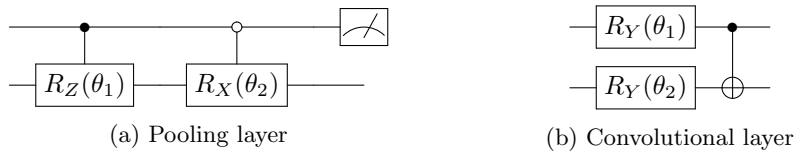

The QCNN consists of a layered architecture as shown in \cite{Hur_2022}, where the number of qubits used by layer decreases with the depth of the circuit. A layer is composed of a convolutional layer, entangling qubits with parameterized gates, and a pooling layer, which also entangles qubits but then reduces the number of qubits by tracing out half of them. Figure \ref{fig:pooling_convolution} gives details about the architecture we used for our convolutional and pooling layers. At the end of the quantum circuit, we measure the last qubit in the Z-basis in order to obtain the expectation value  $\langle \Psi | \sigma_Z | \Psi\rangle$, where $\Psi$ is the quantum state at the end of the quantum circuit.

As mentioned, we have decided to use 8 qubits, so we use a PCA on the image to reduce 8x8=64 features to 8. This reduction is already less critical than reducing a 64x64=4096 image to 8 features, because the correlations made by the PCA do not mix different sections of the PE file. For the encoding of data, we map all input data $x\in\mathbb{R}^n$ in $[0,\frac{\pi}{2}]^n$, and then we apply the encoding map 
\begin{equation}
\begin{split}
    U_\phi: x\mapsto |\phi(x)\rangle = \bigotimes_{i=1}^n\left(\cos(x_i)|0\rangle +\sin(x_i)|1\rangle\right).
\end{split}
\end{equation}

 We build five QCNNs with the same architecture, and then the five QCNNs are trained on their respective dataset, namely images corresponding to one specific section. Scores are available in the Section \ref{Results}, and vary from one section to another. Each QCNN is composed of eight qubits (three layers), and is trained over five epochs. For the optimization method, we use Simultaneous Perturbation for Stochastic Backpropagation (SPSB) \cite{Hoffmann2022}, because it is faster than the parameter shift rule while having comparable results.

 As well as working with a small number of qubits for each QCNN, an interesting advantage is in the interpretation of the results. In addition to the detection score, we are able to see, thanks to the outputs of each QCNN, which section of the binary file is suspicious, and if necessary investigate its content.

\subsection{Training a scoring function}

Once the parameters of each QCNN have been set, we need to find a customized scoring function. One could think about using a majority vote, but this would include a bias that suggests that every section is equally important. However, a missing section can reveal additional information by its absence. In this aim we use the second dataset, and for each binary file, the file is decomposed into images corresponding to the sections present in the binary. The images are then passed through the QCNNs associated with the sections, and each QCNN returns a score. As a reminder, if the section is missing, the score is set to $-1$ for that section.

As an output of this step, we therefore have a vector composed of the five scores, and we want a detection score as a final output. We test various functions and train some algorithms to try and find the most optimal function for calculating this score. For the final version of the algorithm, we keep a XGBoost model, which gave us the best results. The various experiments and results can be found in section \ref{Results}.

\begin{figure}[!h]
    \centering
    \includegraphics[width=1\textwidth]{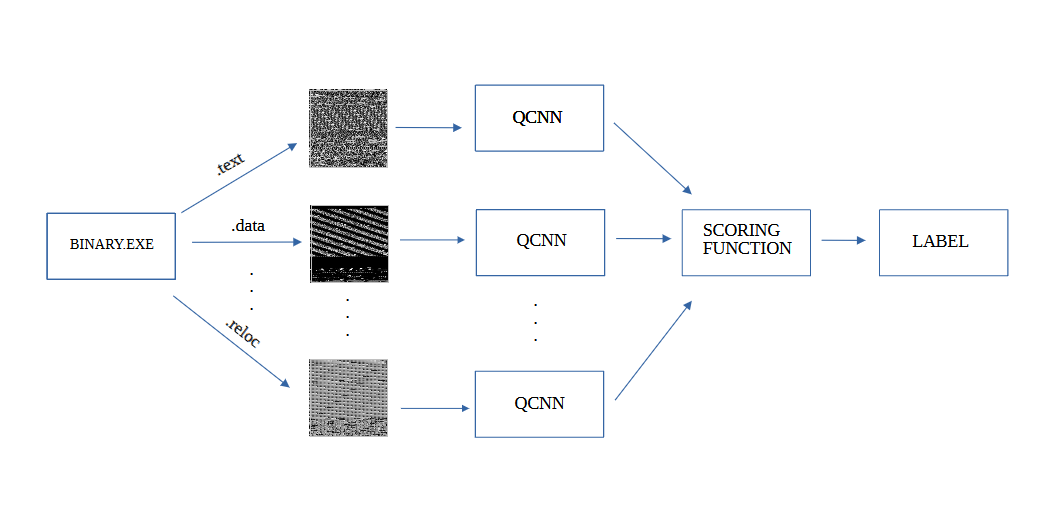}
    \caption{Architecture of our algorithm. We start from a binary file, we decompose it into images of its sections. Each section is used to train a specific QCNN, then we gather the scores of the QCNN in a vector used as input to train a scoring function. Once everything is trained, we get a model which take the binary as input and give its predicted label as output. }
    \label{fig:archi}
\end{figure}

\section{Experiments and results}
\label{Results}

In this section we present the results of our algorithm, which contains three steps. First, we train a QCNN model for each five section of the files. When a section is not present in the file, there is no image corresponding to this section, so the different QCNNs are not trained on exactly the same sizes of dataset, but this guarantees the unbiased property of the training.

Table \ref{tab:section} gathers the training and test accuracy for each section. Since the rate of false-positive and false-negative are very important for our use-case (where sometimes data could be unbalanced) we also compute the F1-score, which gives an additional proof of the performances of our results. We can see that the results by section are not very good, and actually they are comparable to the model where we train the QCNN on the images of the entire files (see Table \ref{tab:full_QCNN}). Note also that the learning seems to be better on some sections, as for example the "rsrc" section when looking at the F1-score, which can tell us about the importance of these sections.  

\begin{table}[!ht]
        \caption{QCNN training results for each section }
        \label{tab:section}
   \centering
    \begin{tabular}{|c|c|c|c|c|c|}
    \hline
      & Train & \multicolumn{2}{c|}{Test} \\
    \hline
     Section & Accuracy & Accuracy & F1-score \\
    \hline
    text & 0.67 & 0.69 &  0.68 \\
    \hline
    data & 0.56 & 0.55 & 0.7 \\
    \hline
    rdata & 0.66 & 0.66 & 0.78 \\
    \hline
    rsrc & 0.65 & 0.63 & 0.72 \\
    \hline
    reloc & 0.69 & 0.7 & 0.67 \\
    \hline
    \end{tabular}
\end{table}

Once the QCNNs are trained, we use them with the second part of the dataset in order to assign to each PE file a vector containing the scores associated to each section. Then we have to train the scoring function on this dataset. In order to identify the good scoring functions for our problem, we try several models, which are presented in Table \ref{tab:scoring_function}. We can see that the XGBoost and the Random Forests (RF) models have the best performances, the RF model being better on the train but comparable on the test. The LightGBM model performs slightly worse than XGBoost, so we do not keep it for the final step. Note that we also try a majority vote, which as expected is not relevant in our use-case.

\begin{table}[!ht]
        \caption{Scoring function training results}
        \label{tab:scoring_function}
   \centering
    \begin{tabular}{|c|c|c|c|c|c|}
    \hline
      & \multicolumn{2}{c|}{Train} & \multicolumn{2}{c|}{Test} \\
    \hline
     Section & Accuracy & F1-score & Accuracy & F1-score \\
    \hline
    XGBoost & 0.89 & 0.90 &  0.84 & 0.85 \\
    \hline
    LGBM & 0.84 & 0.86 & 0.82 & 0.85 \\
    \hline
    RF & 0.99 & 0.99 & 0.85 & 0.86 \\
    \hline
     Maj. Vote &  &  & 0.4 & 0.27 \\
    \hline
    \end{tabular}
\end{table}

Finally, once both the QCNNs and the scoring function are trained, we use the last part of the dataset as a test sample, to evaluate the performances of the whole model. We compare two models in Table \ref{tab:test}, and here we see that the model using XGBoost seems to perform slightly better than the RF model. Note that if we compare this distributed QCNN model to the basic one which use one QCNN for images containing information of the entire files, we have a drastic increase (around 20\%) of both accuracy and F1-score. 

\begin{table}[!ht]
    \begin{center}
        \caption{Testing multi-QCNN with two different scoring function on third dataset}
        \label{tab:test}
        \begin{tabular}{|c|c|c|}
            \hline
             &  Accuracy & F1-score \\ 
             \hline
             RF  &  0.82 & 0.80 \\ 
             \hline
             XGBoost & 0.83 & 0.83 \\
             \hline
        \end{tabular}
    \end{center}
\end{table}

For comparison, we trained a QCNN on the second sub-dataset and then tested it on the third sub-dataset with similar parameters (three layers and five epochs). The results, gathered in Table \ref{tab:full_QCNN}, show that training a model with a PCA on the entire image of file is not a suitable solution. One explanation could be that the PCA has to deal with too many features, and creates correlations between features belonging to different sections.

\begin{table}[!ht]
    \begin{center}
        \caption{Results of QCNN on a complete binary image}
        \label{tab:full_QCNN}
        \begin{tabular}{|c|c|c|}
            \hline
             &  Accuracy & F1-score \\ 
             \hline
             Train  &  0.68 & 0.70 \\ 
             \hline
             Test & 0.60 & 0.66 \\
             \hline
        \end{tabular}
    \end{center}
\end{table}

\section{Conclusion and future work}

The power of our proposed algorithm is that we can explore many different directions to enhance its performances. First we can increase the number of sections taken into account (.idata, .edata or .bss), in order to understand their impact on the final score. This could be done very efficiently as we do not have to re-train the QCNNs on the already studied sections, but just train some new QCNNs on the additional sections and take them into account in the input of the scoring function. We can also investigate what scoring function would be the most relevant for our problem. For example, the use of random forests allows to get information about the sections, and thus to identify which sections are the most likely to help the classification task. Besides, once this information gathered, we could implement a weighted average scoring function to give a percentage of maliciousness of the files, hence giving a more nuanced response to the problem. Finally, we could improve the QCNNs themselves, because we only used one architecture, and some other choices could be made for the convolutional and pooling layers for example. 

In conclusion, we proposed in this work a distributed QCNN model in order to classify malware and benign files. We identified the different sections of these files, and transformed each section into an image thanks to Grayscale method. Then, we selected five specific sections, and trained a QCNN on each section. Once the training done, we gathered the five QCNNs outputs into vectors, that we studied using a scoring function in order to identify the type of each file. Our results for different scoring functions show the efficiency of this model compared to a QCNN classifying images of entire files. Besides, the structure of this model makes it easy to modify, adapt, and allows to get more information about the data, increasing our understanding about malware PE files.

\section*{Acknowledgments}

We want to thank Salomé Quertier for this beautiful drawing of our architecture.

\bibliographystyle{unsrt}
\bibliography{ref}

\end{document}